\documentstyle[preprint,aps,epsf]{revtex}
\begin{document}
\preprint{\font\fortssbx=cmssbx10 scaled \magstep2
\hfill$\vcenter{\hbox{\bf CERN-TH.7508/94}
                \hbox{\bf IFUSP-P 1129}
                \hbox{\bf arch-ive/9411392}
                \hbox{\bf November 1994}
}$}
%
\title{Bounds on Scalar Leptoquarks from Z Physics}
\author{J.\ K.\ Mizukoshi
\footnote{ E-mail: mizukoshi@uspif.if.usp.br (internet) or
47602::MIZUKOSHI (decnet).}, }
\address{Instituto de F\'{\i}sica,  Universidade de S\~ao Paulo, \\
Caixa Postal 20516, 01452-990 S\~ao Paulo, Brazil.}
\author{O.\ J.\ P.\ \'Eboli
\footnote{ Permanent address: Instituto de F\'{\i}sica da
  USP, C.\ P.\ 20.516, 01452-990 S\~ao Paulo, SP, Brazil. E-mail:
  eboli@snfma1.if.usp.br (internet) or 47602::EBOLI (decnet).}
, and
M.\ C.\ Gonzal\'ez-Garc\'{\i}a
\footnote{E-mail: concha@vxcern.cern.ch (internet) or
VXCERN::CONCHA (decnet).}
}
\address{Theory Division,   CERN, \\
CH-1211 Geneva 23, Switzerland.}
\maketitle
\thispagestyle{empty}

\begin{abstract}
  We analyse the constraints on scalar leptoquarks coming from
  radiative corrections to $Z$ physics.  We perform a global fitting
  to the LEP data including the contributions of the most general
  effective Lagrangian for scalar leptoquarks, which exhibits the
  $SU(2)_L \times U(1)_Y$ gauge invariance. We show that the bounds on
  leptoquarks that couple to the top quark are much stronger than the
  ones obtained from low energy experiments.
\end{abstract}

%
%
\medskip
\begin{center}
Submitted to Nuclear Physics B
\end{center}

\section{Introduction}

The Standard Model (SM) of electroweak interactions is extremely
successful in explaining all the available experimental data
\cite{sm}, which is a striking confirmation of the $SU(2)_L\times
U(1)_Y$ invariant interactions involving fermions and gauge bosons.
However, not only has the SM some unpleasant features, like the large
number of free parameters, but it also leaves some questions answered,
such as the reason for the proliferation of fermion generations and
their complex pattern of masses and mixing angles. These problems must
be addressed by models that are intended to be more fundamental than
the SM. A large number of such extensions of the SM predict the
existence of colour triplet particles carrying simultaneously leptonic
and baryonic number, the so-called leptoquarks. Leptoquarks are
present in models that treat quarks and leptons on the same footing,
and consequently quark-lepton transitions are allowed. This class of
models includes composite models \cite{comp}, grand unified theories
\cite{gut}, technicolour models \cite{tech}, and superstring-inspired
models \cite{rizzo}.

Since leptoquarks are an undeniable signal for physics beyond the SM,
there have been several direct searches for them in accelerators.  At
the CERN Large Electron-Positron Collider (LEP), the experiments
established a lower bound $M_{LQ} \gtrsim 45$--$73$ GeV for scalar
leptoquarks \cite{lep}. On the other hand, the search for scalar
leptoquarks decaying into an electron-jet pair in $p\bar{p}$ colliders
constrained their masses to be $M_{LQ} \gtrsim 113$ GeV \cite{ppbar}.
Furthermore, the experiments at the DESY $ep$ collider HERA
\cite{hera} place limits on their masses and couplings, leading to
$M_{LQ} \gtrsim 92-184$ GeV depending on the leptoquark type and
couplings. There have also been many studies of the possibility of
observing leptoquarks in the future $pp$ \cite{fut:pp}, $ep$
\cite{buch,fut:ep}, $e^+e^-$ \cite{fut:ee}, $e\gamma$ \cite{fut:eg},
and $\gamma\gamma$ \cite{fut:gg} colliders.

One can also constrain the masses and couplings of leptoquarks using
the low-energy experiments \cite{leurer,davi}. For energies much
smaller than the leptoquarks mass, they induce two-lepton--two-quark
effective interactions that can give rise to atomic parity violation,
contributions to meson decay, flavour-changing neutral currents
(FCNC), and to meson--anti-meson mixing.  However, only leptoquarks
that couple to first and second generation quarks and leptons are
strongly constrained.

In the present work we study the constraints on scalar leptoquarks
that can be obtained from their contributions to the radiative
corrections to the $Z$ physics. We evaluated the one-loop contribution
due to leptoquarks to all LEP observables and made a global fit in
order to extract the 95\% confidence level limits on the leptoquarks
masses and couplings. The more stringent limits are for leptoquarks
that couple to the top quarks. Therefore, our results turn out to be
complementary to the low energy limits bounds \cite{leurer,davi} since
these constrain more strongly first and second generation leptoquarks.

The outline of this paper is as follows. In Sec.\ \ref{l:eff} we
introduce the effective $SU(2)_L \times U(1)_Y$ invariant effective
Lagrangians that we analysed and we also discuss the existing low
energy constraints.  Section \ref{an:exp} contains the relevant
analytical expressions for the one-loop corrections induced by
leptoquarks.  Our results and respective discussion are shown in Sec.\
\ref{res}, and we summarize our conclusions in Sec.\ \ref{conclusec}.
This paper is supplemented with an appendix that contains the relevant
Feynman rules.


\section{Effective Interaction and Their Low Energy Constraints}
\label{l:eff}

A natural hypothesis for theories beyond the SM is that they exhibit
the gauge symmetry $SU(2)_L \times U(1)_Y$ above the symmetry breaking
scale $v$. Therefore, we imposed this symmetry on the leptoquark
interactions.  In order to avoid strong bounds coming from the proton
lifetime experiments, we required baryon ($B$) and lepton ($L$) number
conservation, which forbids the leptoquarks to couple to diquarks.
The most general effective Lagrangian for leptoquarks satisfying the
above requirements and electric charge and colour conservation is
given by \cite{buch}
\begin{eqnarray}
{\cal L}_{\text{eff}}~ &&=~ {\cal L}_{F=2} ~+~ {\cal L}_{F=0}
\; ,
\nonumber \\
{\cal L}_{F=2}~ &&=~ \left ( g_{\text{1L}}~ \bar{q}^c_L~ i \tau_2~ \ell_L +
g_{\text{1R}}~ \bar{u}^c_R~ e_R \right )~ S_1
+ \tilde{g}_{\text{1R}}~ \bar{d}^c_R ~ e_R ~ \tilde{S}_1
+ g_{3L}~ \bar{q}^c_L~ i \tau_2~\vec{\tau}~ \ell_L \cdot \vec{S}_3
\; ,
\label{lag:fer}
\label{eff} \\
{\cal L}_{F=0}~ &&=~ h_{\text{2L}}~ R_2^T~ \bar{u}_R~ i \tau_2 ~ \ell_L
+ h_{\text{2R}}~ \bar{q}_L  ~ e_R ~  R_2
+ \tilde{h}_{\text{2L}}~ \tilde{R}^T_2~ \bar{d}_R~ i \tau_2~ \ell_L
\; ,
\nonumber
\end{eqnarray}
where $F=3B+L$, $q$ ($\ell$) stands for the left-handed quark (lepton)
doublet, and $u_R$, $d_R$, and $e_R$ are the singlet components of the
fermions. We denote the charge conjugated fermion fields by
$\psi^c=C\bar\psi^T$ and we omitted in (\ref{lag:fer}) the flavour
indices of the couplings to fermions and leptoquarks. The leptoquarks
$S_1$ and $\tilde{S}_1$ are singlets under $SU(2)_L$ while $R_2$ and
$\tilde{R}_2$ are doublets, and $S_3$ is a triplet.  Furthermore, we
assumed in this work that the leptoquarks belonging to a given
$SU(2)_L$ multiplet are degenerate in mass, with their mass denoted by
$M$.

Local invariance under $SU(2)_L \times U(1)_Y$ implies that
leptoquarks also couple to the electroweak gauge bosons. To obtain the
couplings to $W^\pm$, $Z$, and $\gamma$, we substituted $\partial_\mu$
by the electroweak covariant derivative in the leptoquark kinetic
Lagrangian:
\begin{equation}
{\cal L}_{\text{kin}} = \left ( D_\mu \Phi \right )^\dagger D^\mu \Phi
\; ,
\label{lkin}
\end{equation}
with
\begin{equation}
D_\mu \Phi = \left [ \partial_\mu -
i \frac{e}{\sqrt{2} s_W} \left ( W_\mu^+ T^+ + W_\mu^- T^- \right ) -
i e Q_Z Z_\mu
+ i e Q^\gamma A_\mu \right ] \Phi \; ,
\end{equation}
where $\Phi$ stands for the leptoquarks interpolating fields,
$Q^\gamma$ is the electric charge matrix of the leptoquarks, $s_W$ is
the sine of the weak mixing angle, and the $T$'s are the generators of
$SU(2)_L$ for the representation of the leptoquarks. The weak neutral
charge $Q_Z$ is given by
\begin{equation}
Q_Z = \frac{{ T_3 - s_W^2 Q^\gamma }}
{{ s_W c_W}} \; .
\end{equation}
In the last reference of \cite{fut:ee}, for instance, there is a table
with all the quantum numbers for all scalar leptoquarks. In the
Appendix we present the Feynman rules for the interactions defined by
Eqs.\ (\ref{eff}) and (\ref{lkin}).

Low-energy experiments can lead to strong bounds on the couplings and
masses of leptoquarks, due to induced four fermion interactions:

$\bullet$ Leptoquarks can give rise to FCNC processes if they couple
to more than one family of quarks or leptons \cite{shanker,fcnc}. In
order to avoid strong bounds from FCNCs, we assumed that the
leptoquarks couple to a single generation of quarks and a single one
of leptons. However, due to mixing effects on the quark sector, there
is still some amount of FCNC \cite{leurer} and, therefore, leptoquarks
that couple to the first two generations of quarks must comply with
some low-energy bounds \cite{leurer}.

$\bullet$ The analyses of the decays of pseudoscalar mesons, like the
pions, put stringent bounds on leptoquarks unless their coupling is
chiral -- that is, it is either left-handed or right-handed
\cite{shanker}.

$\bullet$ Leptoquarks that couple to the first family of quarks and
leptons are strongly constrained by atomic parity violation
\cite{apv}.  In this case, there is no choice of couplings that avoids
the strong limits.

It is interesting to keep in mind that the low-energy data constrain
the masses of the first generation leptoquarks to be bigger than
$0.5$--$1$ TeV when the coupling constants are equal to the
electromagnetic coupling $e$ \cite{leurer}.


\section{Analytical Expressions}
\label{an:exp}

In this work we employed the on-shell-renormalization scheme, adopting
the conventions of Ref.\ \cite{hollik}. We used as inputs the fermion
masses, $G_F$, $\alpha_{\text{em}}$, and the $Z$ mass, and the
electroweak mixing angle being a derived quantity that is defined
through $\sin^2 \theta_w = s_W^2 \equiv 1 - M^2_W / M^2_Z$. We
evaluated the loops integrals using dimensional regularization
\cite{reg:dim}, since it is a gauge-invariant regularization
procedure, and we adopted the Feynman gauge to perform the
calculations.

Close to the $Z$ resonance, the physics can be summarized by the
effective neutral current
\begin{equation}
J_\mu =  \left ( \sqrt{2} G_\mu M_Z^2 \rho_f
\right )^{1/2} \left [ \left ( I_3^f - 2 Q^f s_W^2 \kappa_f \right )
\gamma_\mu - I_3^f \gamma_\mu \gamma_5 \right ] \; ,
\label{form:nc}
\end{equation}
where $Q^f$ ($I_3^f$) is the fermion electric charge (third component
of weak isospin).  The form factors $\rho_f$ and $\kappa_f$ have universal
contributions, {\em i.e.} independent of the fermion species, as well
as non-universal parts:
\begin{eqnarray}
 \rho_f  &=& 1 + \Delta \rho_{\text{univ}} +
\Delta \rho_{\text{non}} \; , \\
\kappa_f &=& 1 + \Delta \kappa_{\text{univ}} +
\Delta \kappa_{\text{non}} \; .
\end{eqnarray}

\begin{figure}
\epsfxsize=10cm
\begin{center}
\leavevmode \epsfbox{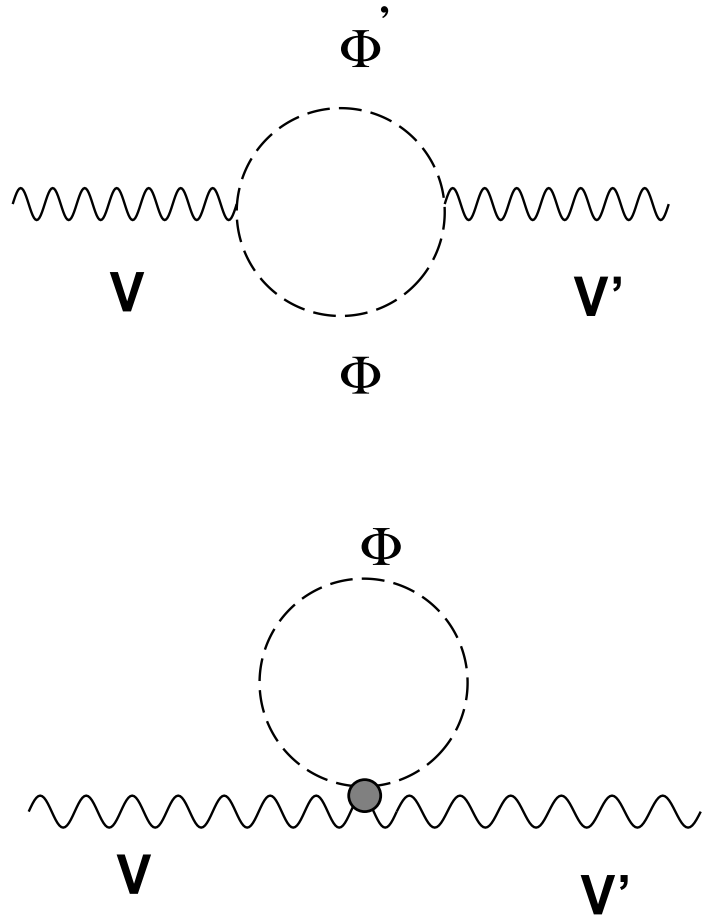}
\end{center}
\caption{Feynman diagrams leading to leptoquark contribution to
vector-boson self-energies.}
\label{prop}
\end{figure}

Leptoquarks can affect the physics at the $Z$ pole through their
contributions to both universal and non-universal corrections. The
universal contributions can be expressed in terms of the
unrenormalized vector boson self-energy ($\Sigma$) as
\begin{eqnarray}
\Delta \rho^{LQ}_{\text{univ}}(s) &=&
-\frac{\Sigma^Z_{LQ}(s)-\Sigma^Z_{LQ}(M_Z^2)}{s-M_Z^2}
+\frac{\Sigma^Z_{LQ}(M_Z^2)}{M_Z^2}
-\frac{\Sigma^W_{LQ}(0)}{M_W^2} - 2 \frac{s_W}{c_W}~
\frac{\Sigma^{\gamma Z}_{LQ}(0)}
{M_Z^2} - \chi_e - \chi_\mu
\; ,\\
\Delta \kappa^{LQ}_{\text{univ}} &=& - \frac{c_W}{s_W}~
\frac{\Sigma^{\gamma Z}_{LQ}(M_Z^2)}{M_Z^2}
- \frac{c_W}{s_W}~
\frac{\Sigma^{\gamma Z}_{LQ}(0)}{M_Z^2}
+\frac{c_W^2}{s_W^2} \left[ \frac{\Sigma_{LQ}^Z(M_Z^2)}{M_Z^2}-
\frac{\Sigma_{LQ}^W(M_W^2)}{M_W^2}\right]
\; ,
\end{eqnarray}
where  the factors $\chi_\ell$ are defined below.  The diagrams with
leptoquark contributions to the self-energies are shown in Fig.\
\ref{prop}.  They can be easily evaluated, yielding
\begin{eqnarray}
{\Sigma}^\gamma_{\text{LQ}}(k^2) &=& - \frac{\alpha_{\text{em}}}{4\pi} N_c
\sum_{j} (Q^\gamma_{j})^2~
{\cal H} \left ( k^2, M^2\right )
\; , \label{sig:g} \\
{\Sigma}^Z_{\text{LQ}}(k^2) &=& - \frac{\alpha_{\text{em}}}
{4\pi } N_c \sum_{j} \left ( Q_Z^{j} \right) ^2
 {\cal H}\left ( k^2, M^2 \right )
\; , \label{sig:z} \\
{\Sigma}^{\gamma Z}_{\text{LQ}} (k^2) &=& \frac{\alpha_{\text{em}}}{4\pi}
N_c \sum_{j} Q^\gamma_{j} Q_Z^{j}
 {\cal H}\left ( k^2, M^2 \right )
\; , \label{sig:gz} \\
{\Sigma}^W_{\text{LQ}}(k^2) &=& - \frac{\alpha_{\text{em}}}
{4\pi s_W^2 } N_c \sum_{j} \left ( T_3^{j} \right )^2
  {\cal H}\left ( k^2, M^2 \right )
\; , \label{sig:w}
\end{eqnarray}
where $N_c = 3$ is the number of colours and the sum is over all
members of the leptoquark multiplet. The function ${\cal H}$ is
defined according to:
\begin{eqnarray}
{\cal H}(k^2, M^2) = &&- \frac{k^2}{3} \Delta_M - \frac{2}{9}k^2
\nonumber \\
&&-  \frac{4 M^2 - k^2}{3} \int^1_0 dx~ \ln \left [
\frac{{ x^2 k^2 - x k^2 + M^2 - i \epsilon}}
{M^2} \right ] \; ,
\end{eqnarray}
with
\begin{equation}
\Delta_M = \frac{2}{4-d} - \gamma_E + \ln(4\pi)  - \ln \left (
\frac{M^2}{\mu^2} \right ) \; ,
\label{delta}
\end{equation}
and $d$ being the number of dimensions. In this section we denote  the
leptoquark masses by capital $M$ with no subindex.

\begin{figure}
\epsfxsize=10cm
\begin{center}
\leavevmode \epsfbox{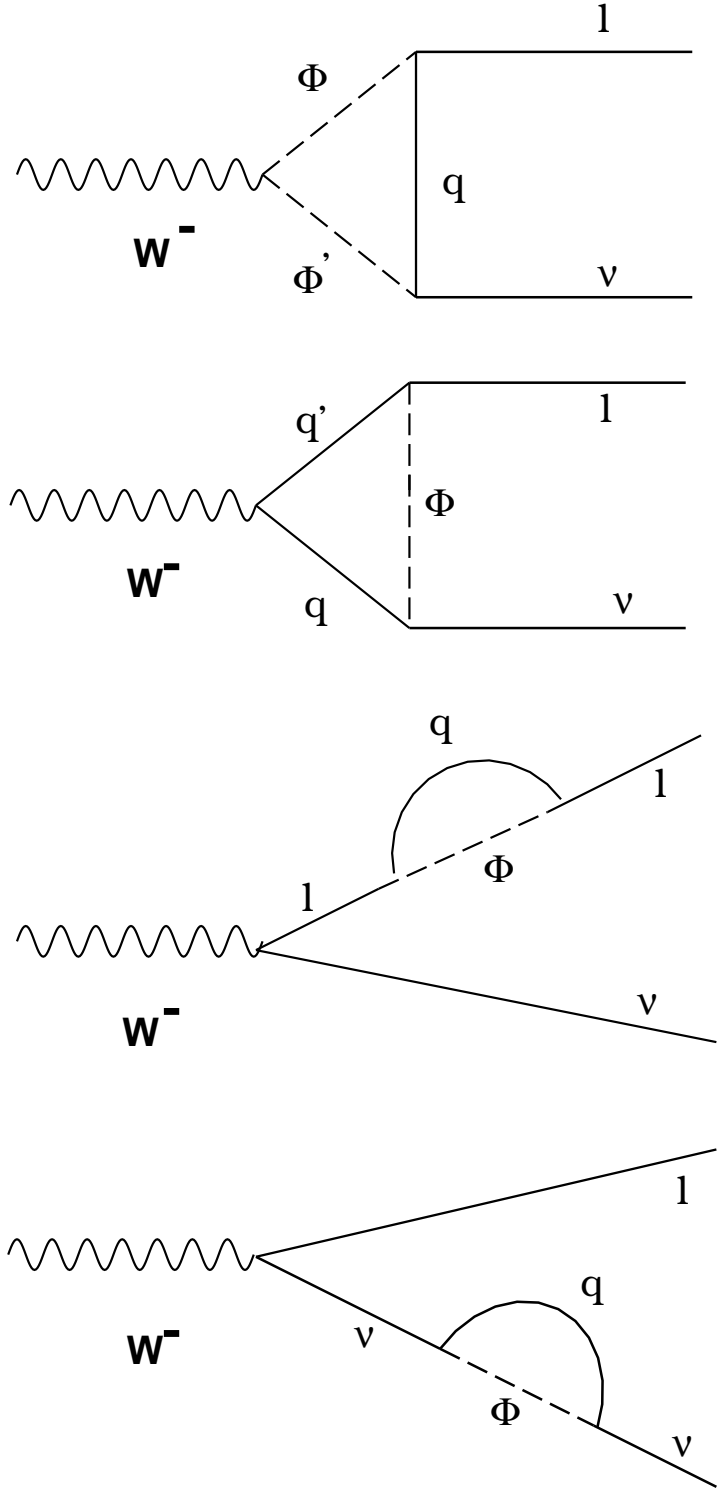}
\end{center}
\caption{ Feynman diagrams leading to leptoquark contribution to
the muon decay. The direction of the lines depends upon
the leptoquark type.}
\label{vertex:w}
\end{figure}

The factors $\chi_\ell$ ($\ell = e$, $\mu$) stem from corrections to
the effective coupling between the $W$ and fermions at low energy.
Leptoquarks modify this coupling through the diagrams shown in Fig.\
\ref{vertex:w}, inducing a contribution that we parametrize as
\begin{equation}
i \frac{e}{\sqrt{2} s_W}~ \chi_\ell~ \gamma_\mu P_L \; ,
\end{equation}
where $P_L$ ($P_R$) is the left-handed (right-handed) projector and
$\ell$ stands for the lepton flavour.  Since this correction modifies
the muon decay, it contributes to $\Delta r$, and consequently, to
$\Delta \rho_{\text{univ}}$. Leptoquarks with right-handed couplings,
as well as the $F=0$ ones, do not contribute to $\chi_\ell$, while the
evaluation of the diagrams of Fig.\ \ref{vertex:w}, for left-handed
leptoquarks in the $F=2$ sector, leads to
\begin{equation}
\begin{array}{ll}
\chi_\ell = & {\displaystyle \lim_{p_W^2 \rightarrow 0}
\frac{g^2_{\text{LQ,L}}}{16 \pi^2} }N_c  \left\{
\frac{1}{2} {\displaystyle \sum_{\text{$j$,q}}}
{M^{j}_{\ell q}}^\dagger M^{j}_{q \ell}
B_1(0, m_q^2, M^2)
+ \frac{1}{2} {\displaystyle
\sum_{\text{$j$,$q^\prime$}}} {M^{j}_{\nu q^\prime}}^\dagger
M^{j}_{q^\prime \nu} B_1(0, m_{q^\prime}^2, M^2)
\right. \\
 & {\displaystyle \sum_{\text{q, $q^\prime$, $j$}}}
{M^{j}_{\ell q}}^\dagger
M^{j}_{q^\prime \nu} V_{q q^\prime} \left [ (2-d) C_{00}(0, p_W^2, 0,
M^2, m_{q^\prime}^2, m_q^2 ) + p_W^2 C_{12} (0, p_W^2, 0,
M^2, m_{q^\prime}^2, m_q^2 ) \right ]  \\
 &\left.  - 2 {\displaystyle \sum_{\text{q, $j$, $j^\prime$}}}
T^-_{j, j^\prime}
{M^{j^\prime}_{\ell q}}^\dagger M^{j}_{q\nu}
C_{00} ( 0, p_W^2, 0, m_q^2, M^2, M^2 ) \right\} \; ,
\end{array}
\label{w:ff2}
\end{equation}
where $j$ stands for the leptoquarks belonging to a given multiplet
and $V_{q q^\prime}$ is the Cabibbo-Kobayashi-Maskawa mixing matrix
for the quarks, which, for simplicity, we will take to be the unit
matrix\footnote{In general, this expression is divergent and requires
  a renormalization of the elements of the CKM matrix.}. The functions
$B_1$, $C_0$, $C_{00}$, and $C_{12}$ are the Passarino-Veltman
functions \cite{passa}.  It is interesting to notice that for
degenerate massless quarks the above expression vanishes, and none of
the leptoquarks contribute to $\chi_\ell$. On the other hand, for the
leptoquarks coupling to the third family (neglecting the bottom quark
mass), we have that
\begin{equation}
\chi_\ell = \frac{g_{1L}^2}{64 \pi^2} N_c \left \{
 \frac{m_{\text{top}}^2}{  M^2 - m_{\text{top}}^2}
{}~+~ \frac{m_{\text{top}}^4 - 2 m_{\text{top}}^2 M^2 }
{ ( M^2 - m_{\text{top}}^2 )^2 }~ \ln \left (
\frac{M^2}{m_{\text{top}}^2} \right )
\right \}   \; ,
\label{chi:s1}
\end{equation}
for the $S_{1L}$ leptoquark, while the $S_{3L}$
contribution can be obtained from (\ref{chi:s1}) by the
substitution $g_{1L}^2 \Rightarrow - g_{3L}^2$.

\begin{figure}
\epsfxsize=10cm
\begin{center}
\leavevmode \epsfbox{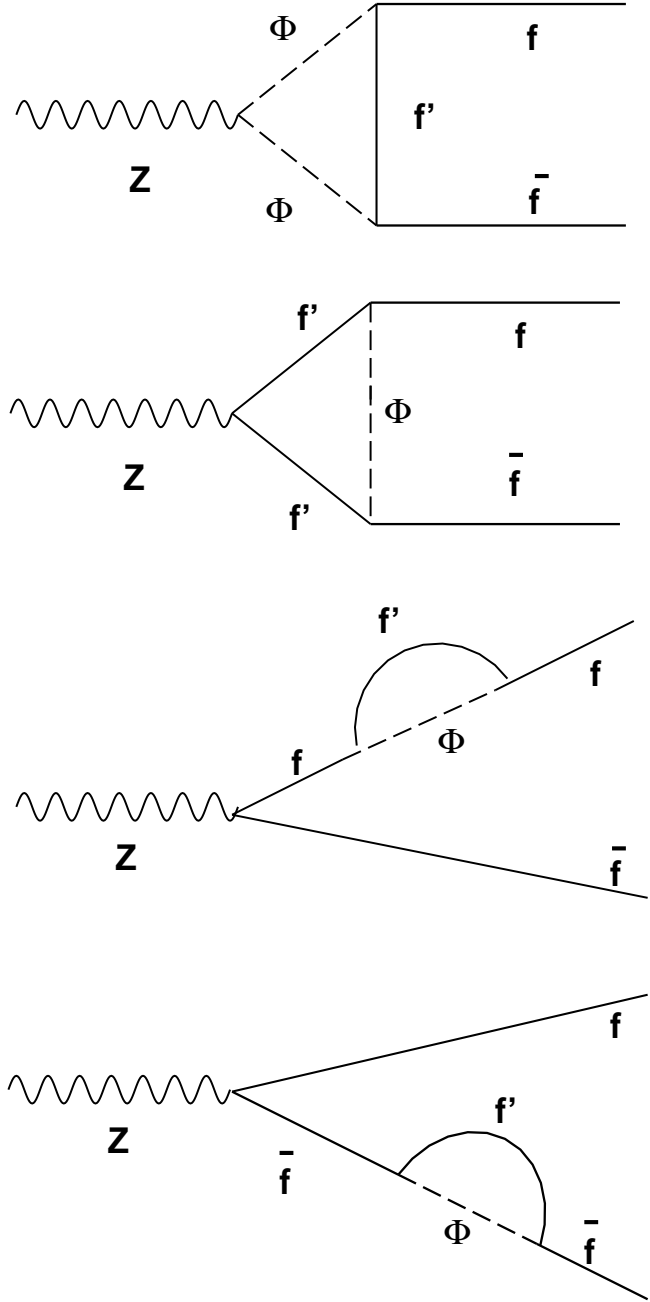}
\end{center}
\caption{Feynman diagrams leading to leptoquark contribution to
the vertex $Z f \bar{f}$. The direction of the lines depends upon
the leptoquark type.}
\label{vertex:z}
\end{figure}

Corrections to the vertex $Z f \bar{f}$ give rise to non-universal
contributions to $\rho_f$ and $\kappa_f$.  Leptoquarks affect these
couplings of the $Z$ through the diagrams given in Fig.\
\ref{vertex:z} whose results we parametrize as
\begin{equation}
i \frac{e}{2 s_W c_W} \left [ \gamma_\mu F_{VLQ}^{Zf} - \gamma_\mu \gamma_5
F_{ALQ}^{Zf} + I_3^f \gamma_\mu (1 - \gamma_5) \frac{c_W}{s_W} ~
\frac{\Sigma^{\gamma Z}_{LQ}(0)}{M_Z^2} \right ] \; ,
\end{equation}
where for leptons ($\ell$) and leptoquarks with $F=2$
\begin{equation}
\begin{array}{ll}
F^\ell_{VLQ}=& \pm F^\ell_{ALQ}=  \frac{g_{LQ,X}^2}{32 \pi^2} N_c
{\displaystyle \sum_{j, q} }
{M^{j}_{\ell q}}^\dagger M^{j}_{q\ell} \\
& \left\{ \frac{g^q_X}{2}
- s_W c_W Q_Z^{j}- \left (g_X^q + 2 s_W c_W Q_Z^{j} \right )~
\frac{M^2 - m_q^2}{M_Z^2}
\left [ - \frac{1}{2} \ln \left ( \frac{M^2}{m_q^2} \right )
+ \bar{B_0} ( 0, m_q^2,M^2 ) \right ] \right. \\
& + 2 s_W c_W Q_Z^{j} \frac{M^2 - m_q^2 - \frac{1}{2} M_Z^2}{M_Z^2}
\left [ - \ln \left ( \frac{M^2}{m_q^2} \right ) + \bar{B_0}
( M_Z^2, M^2, M^2) \right ]  \\
&+ g_X^q \frac{M^2-m_q^2 - \frac{1}{2} M_Z^2}{M_Z^2} \bar{B_0}
(M_Z^2, m_q^2, m_q^2 ) + g^{\ell}_X \bar{B_1} (0, m_q^2, M^2) \\
 & + \left [ g_{-X}^q m_q^2 + g_X^q \frac{(M^2-m_q^2)^2}{M_Z^2}
\right ] C_0 (0, M_Z^2, 0, M^2, m_q^2, m_q^2 ) \\
&\left. - 2 s_W c_W Q_Z^{j} \frac{(M^2-m_q^2)^2 + m_q^2 M_Z^2}{M_Z^2}
C_0 (0, M_Z^2, 0, m_q^2, M^2, M^2) \right\} \; ,
\end{array}
\label{z:ll}
\end{equation}
where the $+$ $(-)$ corresponds to left- (right-) handed leptoquarks and
$g_{L/R}^f = v^f \mp a^f$, the neutral current couplings being
$a_f = I_3^f$ and $v_f = I_3^f - 2 Q^f s_W^2$. We used the convention
$X=L,R$ and $-L=R$ ($-R=L$). We also defined
\begin{eqnarray}
B_0 (k^2, M^2, {M^\prime}^2) &\equiv& \frac{1}{2}\Delta_M+
\frac{1}{2} \Delta_{M'} + \bar{B_0}
(k^2, M^2, {M^\prime}^2 )
\; , \\
B_1 (k^2, M^2, {M^\prime}^2) &\equiv& - \frac{1}{2} \Delta_M + \bar{B_1}
(k^2, M^2, {M^\prime}^2)
\; ,
\end{eqnarray}
with $\Delta_M$ given by Eq.\ (\ref{delta}). From this last expression
we can obtain the effect of $F=2$ leptoquarks on the vertex $Z q
\bar{q}$ simply by the change $\ell \Leftrightarrow q$.  Moreover, we
can also employ the expression (\ref{z:ll}) to $F=0$ leptoquarks
provided we substitute $g_{LQ,X} \Rightarrow h_{LQ,X}$ and $g^q_{\pm
  X} \Rightarrow - g^q_{\mp X}$.

With all this we have
\begin{eqnarray}
\Delta \rho^{LQ}_{\text{non}} &=& \frac{F_{ALQ}^{Zf}}{a_f}(M_Z^2)
\; , \\
\Delta \kappa^{LQ}_{\text{non}} &=& - \frac{1}{2 s_W^2  Q^f} \left [
F_{VLQ}^{Zf}(M_Z^2) - \frac{v_f}{a_f}~ F_{ALQ}^{Zf}(M_Z^2)
\right ]
\; .
\end{eqnarray}

One very interesting property of the general leptoquark interactions
that we are analysing is that all the physical observables are
rendered finite by using the same counter-terms as appear in the SM
calculations \cite{hollik}. For instance, starting from the
unrenormalized self-energies (\ref{sig:g}--\ref{sig:w}) and the mass
and wave-function counter-terms we obtain finite expression for the
two-point functions of vector bosons. Moreover, the contributions to
the vertex functions $Z f \bar{f}$ and $W f \bar{f^\prime}$ are
finite, as can be seen from the explicit expressions (\ref{w:ff2}) and
(\ref{z:ll}).

In order to check the consistency of our calculations, we analysed the
effect of leptoquarks to the $\gamma f \bar{f}$ vertex at zero
momentum, which is used as one of the renormalization conditions in
the on-shell scheme. This vertex function can be obtained from
(\ref{z:ll}) by the substitutions $Q_Z^j \Rightarrow -Q_j^\gamma$,
$g^f_X = Q_f$, $e/2s_Wc_W \Rightarrow -e$, and $M_Z^2 \Rightarrow
k^2$, with $k^2$ being the squared four momentum of the photon. It
turns out that the leptoquark contribution to the vertex function
$\gamma f \bar{f}$ not only is finite but also vanishes at $k^2=0$ for
all fermion species.  Therefore, our expressions for the different
leptoquark contributions satisfy the appropriate QED Ward identities
\cite{ward}, and leave the fermion electric charges unchanged.
Moreover, we also verified explicitly that the leptoquarks decouple in
the limit of large $M$.


\section{Results and Discussion}
\label{res}

The above expressions for the radiative corrections to $Z$ physics due
to leptoquarks are valid for couplings and masses of any leptoquark.
For the sake of simplicity, we assumed that the leptoquarks couple to
leptons and quarks of the same family.  In the conclusions, we comment
on how we can extend our results to other cases.

In order to gain some insight on which corrections are the most
relevant, let us begin our analyses by studying just the oblique
corrections \cite{obli}, which can be parametrized in terms of the
variables $\epsilon_1$, $\epsilon_2$, and $\epsilon_3$. The parameter
$\epsilon_1$ vanishes since it is proportional to $\Delta \rho(0)$ and
we assumed the multiplets to be degenerate. The parameter $\epsilon_3$
also vanishes, while
\begin{eqnarray}
\epsilon_2 = \frac{2 N_c}{\pi} \sum_j \left ( T_3^j \right )^2~~
&&\left\{- \left (\frac{1}{6} - \frac{2}{3}~\frac{M^2}{M_W^2} \right )
\bar{B_0}(M^2_W, M^2, M^2)\right.
\nonumber \\
&&
+\left. \left (\frac{1}{6} - \frac{2}{3}~\frac{M^2}{M_Z^2} \right )
\bar{B_0}(M^2_Z, M^2, M^2) \right\}
\; ,
\end{eqnarray}
where the sum is over all members of a given multiplet.  Leptoquarks
that are singlets under $SU(2)_L$ also lead to $\epsilon_2=0$.  Notice
that the above results depend only upon the interaction of leptoquarks
with the gauge bosons.  Imposing that the leptoquark contribution to
$\epsilon_2$ must be within the limits allowed by the LEP data
\cite{altarellinew}, we find out that the constraints coming from
oblique corrections are less restrictive than the available
experimental limits \cite{lep,ppbar,hera}. Therefore, the contribution
of the leptoquarks to the oblique parameters are very small and do not
lead to new bounds.

We then performed a global fit to all LEP data including both
universal and non-universal contributions.  In Table \ref{LEPdata} we
show the most recent combined results of the four LEP experiments.
These results are not statistically independent and the correlation
matrix can be found in \cite{sm}.  We can express the theoretical
predictions to these observables in terms of $\kappa^f$, $\rho^f$, and
$\Delta r$, with the SM contributions being obtained from the program
ZFITTER \cite{zfit}.  In order to perform the global fit we
constructed the $\chi^2$ function and minimized it using the package
MINUIT. In our fit we used five parameters, three from the SM
($m_{\text{top}}$, $M_H$, and $\alpha_s(M_Z^2)$) and two new ones: $M$
and $g_{LQ}$.  We present our results as 95\% CL lower limits in the
leptoquark mass and study the dependence of these limits upon all the
other parameters.

\begin{figure}
\epsfxsize=15cm
\begin{center}
\leavevmode \epsfbox{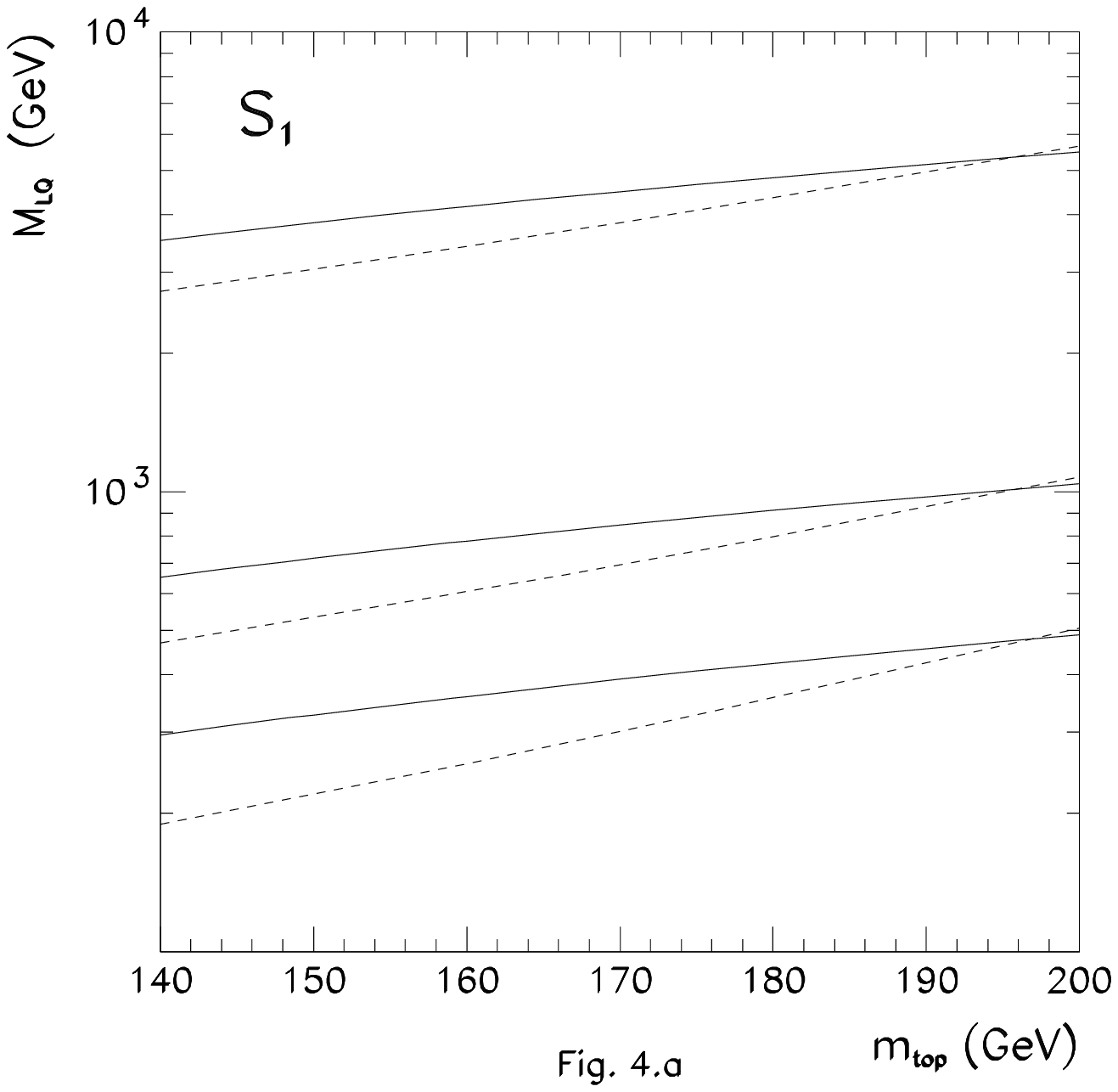}
\end{center}
\epsfxsize=15cm
\begin{center}
\leavevmode \epsfbox{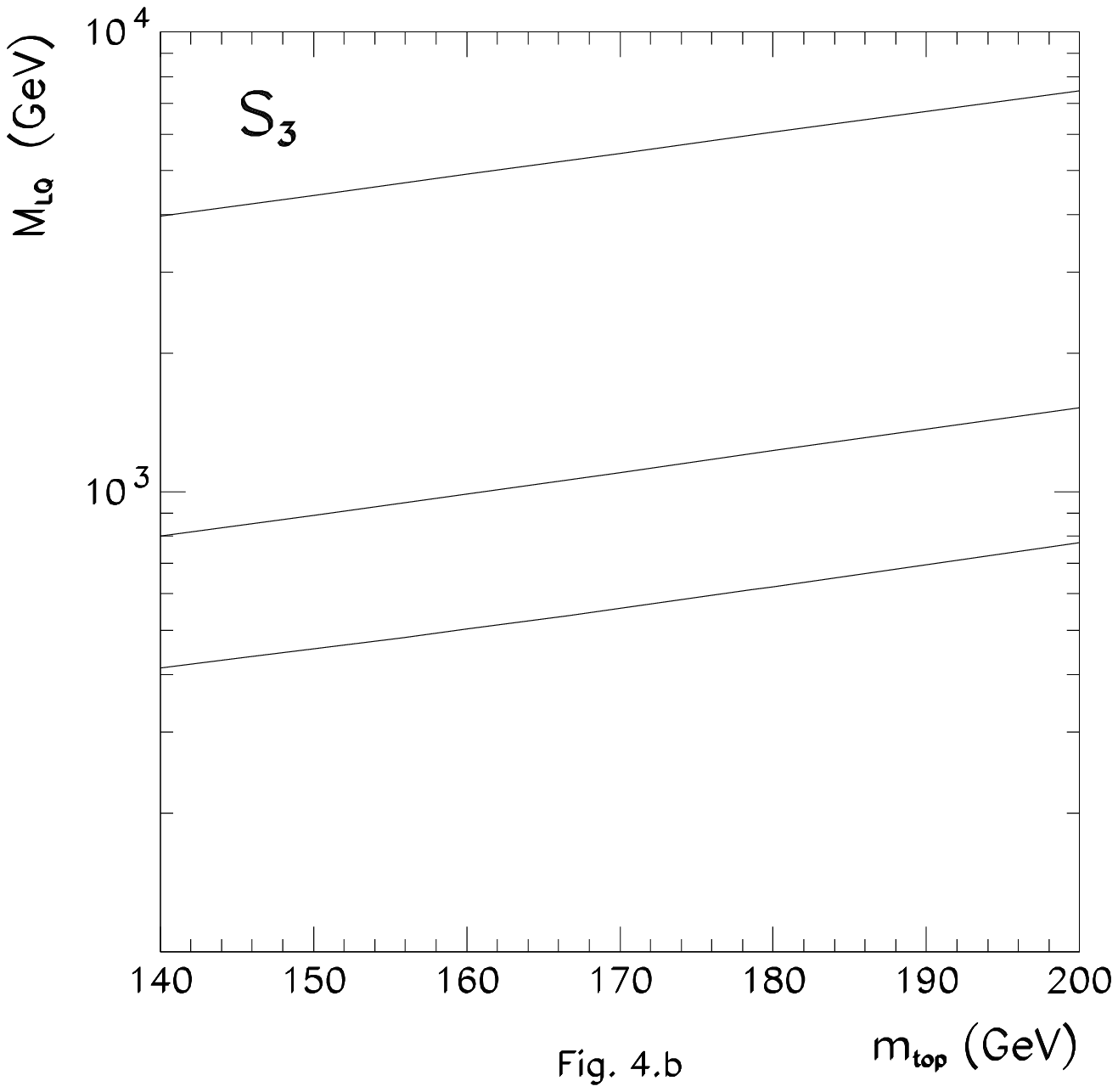}
\end{center}
\epsfxsize=15cm
\begin{center}
\leavevmode \epsfbox{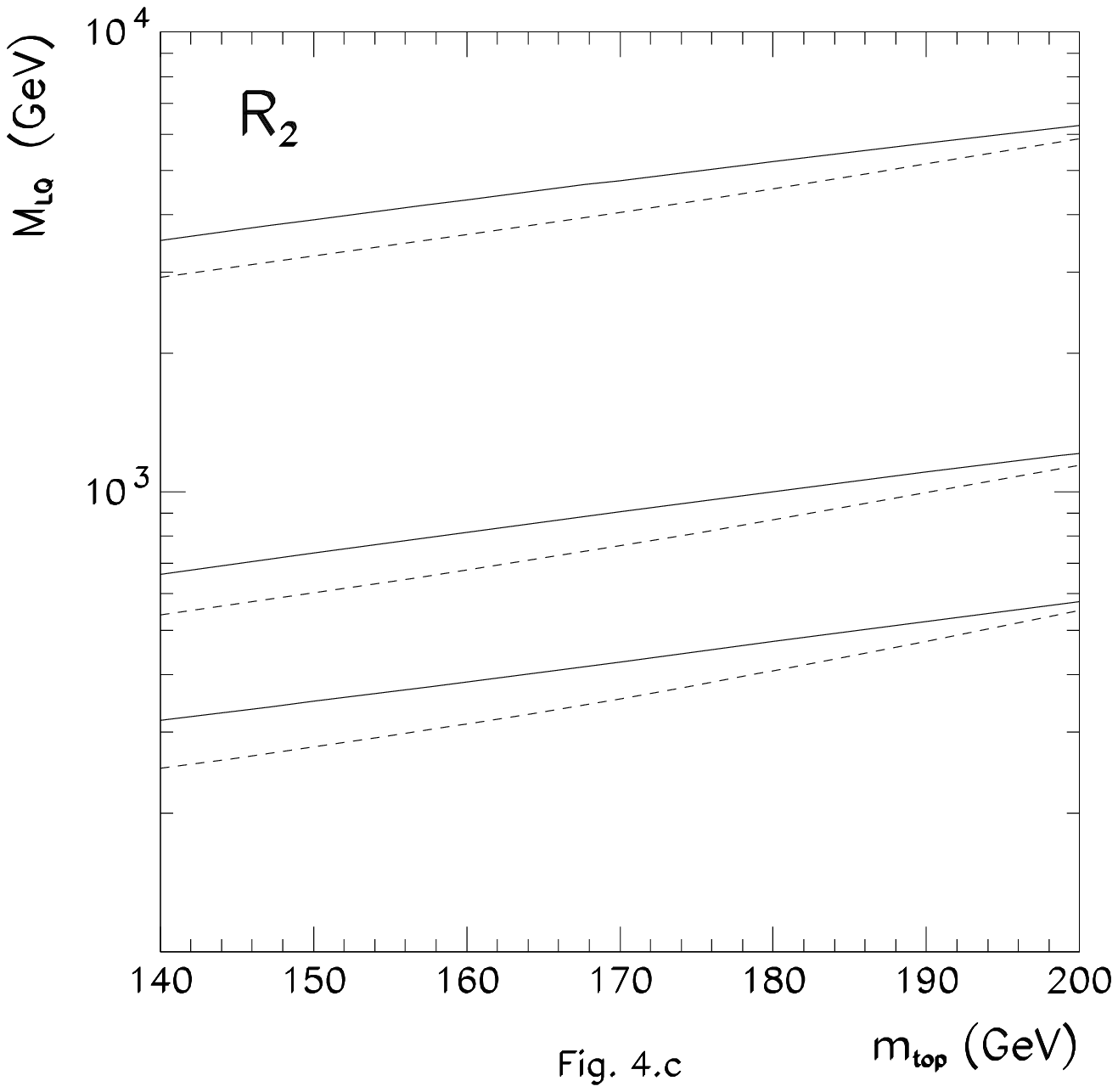}
\end{center}
\caption{Bounds (95\% CL) on the leptoquark masses as a function of
  $m_{\text{top}}$ for $S_1$ (a), $S_3$ (b), and $R_2$ (c).  The solid
  (dashed) lines stand for left-handed (right-handed) leptoquarks,
  while upper (medium, lower) lines correspond to a coupling
  $g_{LQ}=\protect\sqrt{4\pi}$ ($1$, $e/s_W$). }
\label{fit1}
\end{figure}

The parameter of the SM that most strongly affects our results is the
top mass ($m_{\text{top}}$), as expected. For this reason, Figs.\
\ref{fit1} exhibit the 95\% CL limits obtained for a third generation
leptoquark as a function of $m_{\text{top}}$ for several values of the
coupling constants $g_{LQ}$ ($=\sqrt{4\pi}$, 1, and $e/s_W$).  In
these figures, we fixed the value of $M_H=300$ GeV and
$\alpha_s(M_Z^2)=0.126$, which are the best values obtained from a fit
in the framework of the SM \cite{sm}.  We can see from these figures
that the limits on the leptoquarks $S_1$, $S_3$, and $R_2$ become
stronger as $m_{\text{top}}$ increases.  This result is analogous to
the SM result for the radiative corrections to $Z\rightarrow b
\bar{b}$ \cite{pich}, where there is an enhancement by powers of the
top quark mass, as can be seen, for instance, in Eq.\ (\ref{chi:s1}).
We can also learn from these figures that the limits are better for
left-handed leptoquarks than for right-handed ones, given a leptoquark
type.

The contributions from $\tilde R_2$ and $\tilde S_1$ are not enhanced
by powers of the top quark mass since these leptoquarks do not couple
directly to up-type quarks. Therefore, the limits are much weaker,
depending on $m_{\text{top}}$ only through the SM contribution, and
the bounds for these leptoquarks are worse than the present discovery
limits unless they are strongly coupled ($g_{LQ}^2 = 4 \pi$).
Therefore, we do not plot the bounds for these cases.  Moreover, the
limits on first and second generation leptoquarks are also
uninteresting for the same reason.  Nevertheless, if we allow
leptoquarks to mix the third generation of quarks with leptons of
another generation the bounds obtained are basically the same as the
ones discussed above\footnote{In the case of first generation leptons,
  we must also add a tree level $t$-channel leptoquark exchange to
  some observables.}, since the main contribution to the constraints
comes from the $Z$ widths.

Let us finally comment on the dependence of our bounds on other SM
parameters besides $m_{\text{top}}$. For a fixed $m_{\text{top}}$, the
dependence on $\alpha_s(M_Z^2)$ and $M_H$ is rather weak, but it shows
the pattern that the limits are more stringent as $\alpha_s(M_Z^2)$
increases and $M_H$ decreases. For instance, the limits vary by about
$10 \%$ when $\alpha_s(M_Z^2)$ is in the interval
$0.12<\alpha_s(M_Z^2)<0.13$ and by about $25 \%$ for $60<M_H<1000$ GeV.

\section{Summary and Conclusions}
\label{conclusec}

We summarize our results in Table \ref{res:top}, where we assumed a
top mass of $175$ GeV\footnote{Our results for $R_{2L}$ are in
  qualitative agreement with those of Ref.\ \cite{ellis}.}.  Our
analyses show that the LEP data constrain more strongly the
leptoquarks that couple to the top quark. Since the most important
ingredients in the bounds are the widths of the $Z$, we can conclude
that our results should also give a good estimate for the cases where
the leptoquarks couple to quarks and leptons from different families.
It is interesting to notice that the constraints on leptoquarks coming
from LEP data are complementary to the low-energy ones, since these
are more stringent for leptoquarks that couple to the first two
families, while the former are stronger when the coupling is to the
top quark.

The bounds on scalars leptoquarks coming from low-energy and $Z$
physics exclude large regions of the parameter space where the new
collider experiments could search for these particles, however, not
all of it \cite{fut:pp,fut:ep,fut:ee,fut:eg,fut:gg}.  Nevertheless, we
should keep in mind that nothing substitutes the direct observation.


\begin{center}
{\bf ACKNOWLEDGEMENTS}
\end{center}

We want to thank D. Bardine and A. Olchevsky for providing us with the
latest version of ZFITTER.  O.J.P.E. is grateful to the CERN Theory
Division for its kind hospitality. This work was supported by Conselho
Nacional de Desenvolvimento Cient\'{\i}fico e Tecnol\'ogico
(CNPq-Brazil) and by Funda\c{c}\~ao de Amparo \`a Pesquisa do Estado
de S\~ao Paulo (FAPESP).

\newpage


\appendix
\section*{Feynman rules for scalar leptoquarks}
In this section we denote by $\Phi$ the scalar leptoquark multiplet that
is a vector in the weak isospin space.
The Feynman rules for the gauge couplings of scalar leptoquarks are
\epsfxsize=12cm
\begin{center}
\leavevmode\epsfbox{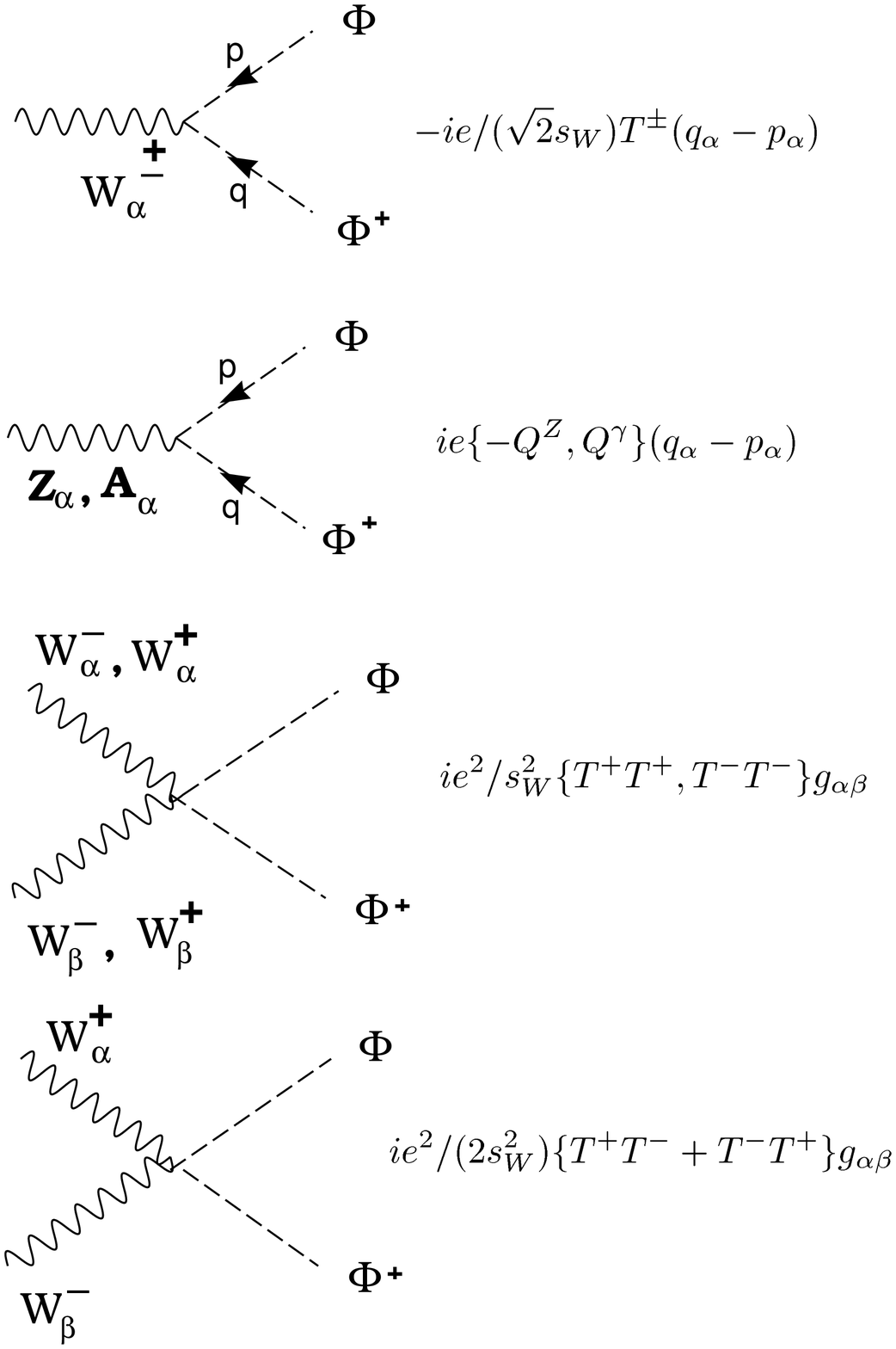}
\end{center}
\epsfxsize=13cm
\begin{center}
\leavevmode\epsfbox{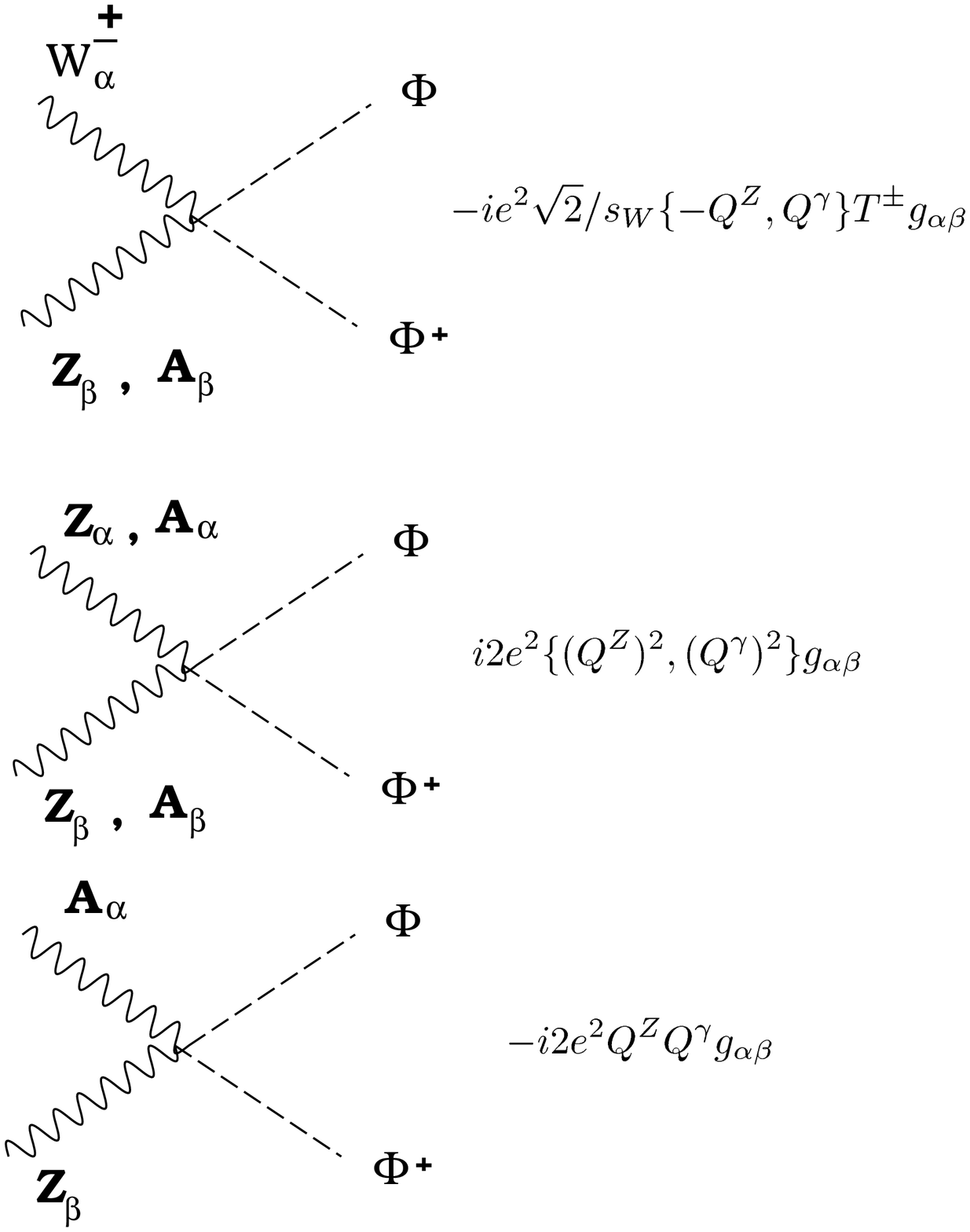}
\end{center}
\newpage
The couplings to fermions are different for $F=2$ and $F=0$ leptoquarks. They
can be parametrized as:
\epsfxsize=11cm
\begin{center}
\leavevmode\epsfbox{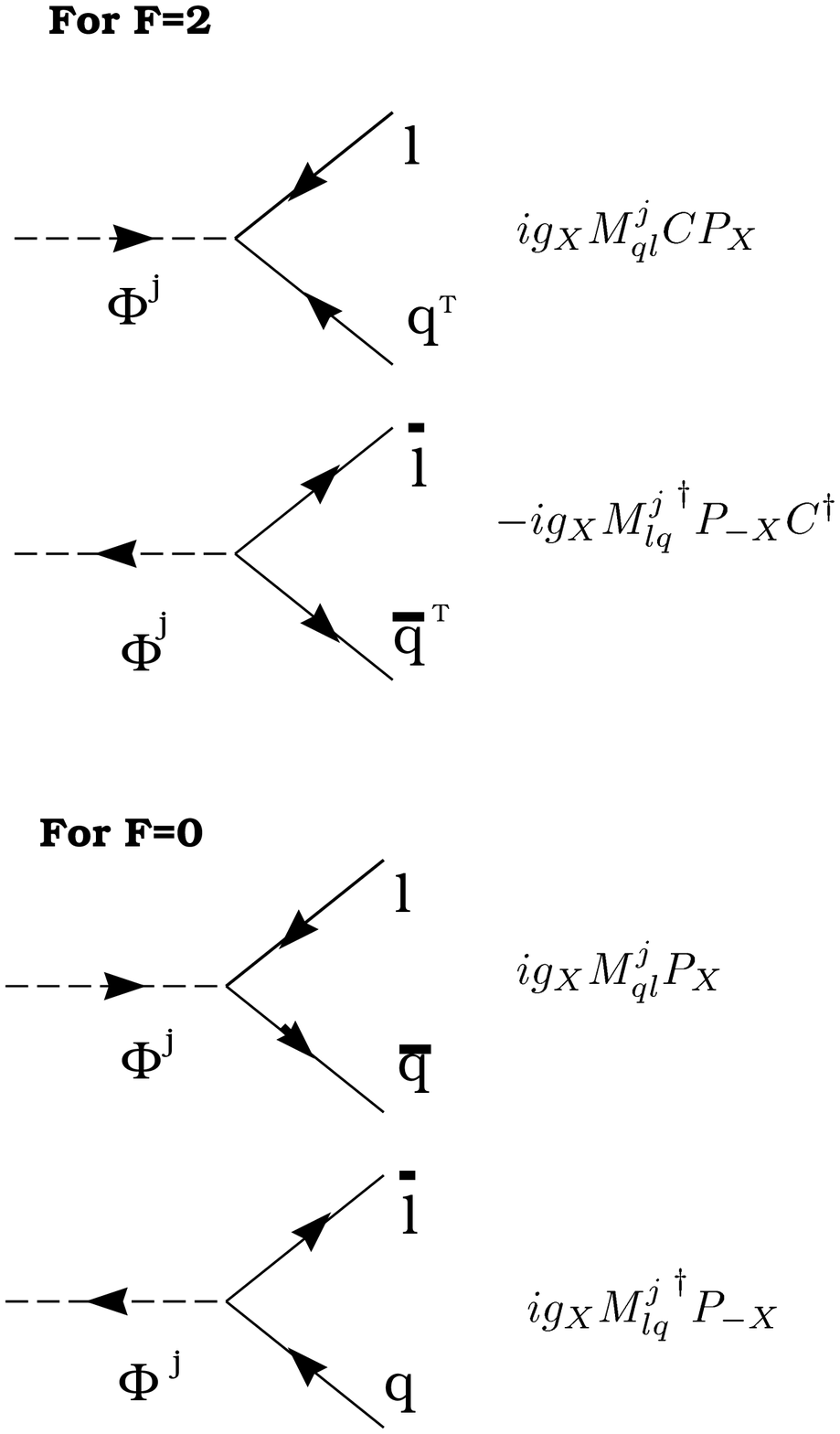}
\end{center}
\newpage
\noindent
where for $S_1$
\begin{equation}
\begin{array}{ll}
M_{L}=\left(\begin{array}{cc} 0 & 1\\ -1 & 0\end{array} \right) \; ,
 \;\;\;\;\;\;\;
&
M_{R}=\left(\begin{array}{cc} 0 & 1\\ 0 & 0\end{array} \right)  \\
\end{array} \; ,
\end{equation}
while for $\tilde S_1$
\begin{equation}
M_{R}=\left(\begin{array}{cc} 0 & 0\\ 0 & 1\end{array} \right) \; .
\end{equation}
The triplet $S_3$ in the diagonal basis $S_3=(S_3^+,S_3^0,S_3^-)$
leads to
\begin{equation}
\begin{array}{lll}
M^+_{L}=\sqrt{2}\left(\begin{array}{cc} 0 & 0\\ 0 & -1\end{array} \right)
  \;\;\;\;\;\;\;
&
M^0_{L}=\left(\begin{array}{cc} 0 & -1\\ -1 & 0\end{array} \right)
\;\;\;\;\;\;\;
&
M^-_{L}=\sqrt{2} \left(\begin{array}{cc} 1 & 0\\ 0 & 0\end{array} \right)
   \;\;\;\;\;\;\;
\end{array}
\; ,
\end{equation}
and the doublet $R_2=(R_2^{1/2},R_2^{-1/2})$  is associated to
\begin{equation}
\begin{array}{ll}
M^+_{L}=\left(\begin{array}{cc} 0 & 1\\ 0 & 0\end{array} \right)
 \;\;\;\;\;\;\;
&
M^-_{L}=\left(\begin{array}{cc} -1 & 0\\ 0 & 0\end{array} \right)
\; , \\[+1.cm]
M^+_{R}=\left(\begin{array}{cc} 0 & 1\\ 0 & 0\end{array} \right)
\;\;\;\;\;\;\;
&
M^-_{R}=\left(\begin{array}{cc} 0 & 0\\ 0 & 1\end{array} \right)\; .  \\
\end{array}
\end{equation}
For $\tilde R_2$ we have
\begin{equation}
\begin{array}{ll}
M^+_{L}=\left(\begin{array}{cc} 0 & 0\\ 0 & 1\end{array} \right)
 \;\;\;\;\;\;\;
&
M^-_{L}=\left(\begin{array}{cc} 0 & 0\\ -1 & 0\end{array} \right)
\end{array}
\; .
\end{equation}




\protect


\protect


\protect

\protect


\begin{table}
\caption{LEP data}
\label{LEPdata}
\begin{displaymath}
\begin{array}{|l|l|}
\hline
\hline
\mbox{Quantity} & \mbox{Experimental value} \\ \hline
M_Z \mbox{[GeV]} & 91.1888 \pm 0.0044 \\
\Gamma_Z \mbox{[GeV]} & 2.4974 \pm 0.0038 \\
\sigma_{\rm had}^0  \mbox{[nb]} & 41.49 \pm 0.12\\
R_e = \frac{\Gamma({\rm had})}{\Gamma(e^+ e^-)} & 20.850 \pm 0.067 \\
R_\mu = \frac{\Gamma({\rm had})}{\Gamma(\mu^+ \mu^-)} & 20.824 \pm 0.059 \\
R_\tau = \frac{\Gamma({\rm had})}
{\Gamma(\mu^+ \mu^-)} & 20.749 \pm 0.070 \\
A_{FB}^{0e} & 0.0156 \pm 0.0034 \\
A_{FB}^{0\mu} & 0.041 \pm 0.0021 \\
A_{FB}^{0\tau} & 0.0228 \pm 0.0026 \\
A_{\tau}^0 & 0.143 \pm 0.010 \\
A_e^0  & 0.135 \pm 0.011 \\
R_b = \frac{\Gamma(b \bar{b})}{ \Gamma({\rm had})} &0.2202 \pm 0.0020\\
R_c = \frac{\Gamma(c\bar{c}) }{\Gamma({\rm had})} & 0.1583 \pm 0.0098\\
A_{FB}^{0b} & 0.0967 \pm 0.0038  \\
A_{FB}^{0c} & 0.0760 \pm 0.0091  \\
\hline
\hline
\end{array}
\end{displaymath}
\end{table}

\widetext
\begin{table}
\caption{ Lower limits (95 \% CL) for the mass of third generation
leptoquarks in
GeV for different values of the couplings, assuming $m_{\text{top}} = 175$
GeV, $\alpha_s(M_Z^2) = 0.126$, and $M_H = 300$ GeV.}
\label{res:top}
\begin{tabular}{|c|c|c|c|c|c|c|c|}
g & $S_1^R$     &   $S_1^L$    &   $S_3$      &   $R_2^R$    &   $R_2^L$
& $\tilde{S}_1^R$ & $\tilde{R}_2^L$ \\
\tableline
$\protect\sqrt{4\pi}$ & 4000 & 5000 & 6000 & 4000 & 5000 & 280 & 400\\
1 & 750 & 900 & 1100 & 850 & 1000 & --- & ---\\
${\displaystyle \frac{e}{s_W}}$ & 350 & 400 & 600 & 400 & 500 & --- & ---
\end{tabular}
\end{table}

\end{document}